\documentclass[aps,prl,10pt,twocolumn,floatfix]{revtex4-2}
\usepackage{amssymb}
\usepackage{natbib}
\usepackage{graphicx}
\usepackage{amsmath}
\usepackage[bookmarks = false,colorlinks = true,allcolors = blue]{hyperref}
\usepackage{bm}
\usepackage[all]{hypcap}
\usepackage{graphicx}
\usepackage{colortbl}
\usepackage{booktabs}
\usepackage{enumerate}
\usepackage{enumitem}
\usepackage{braket}
\usepackage{mathrsfs}
\usepackage{multirow}
\usepackage{upgreek}
\usepackage{textcomp}
\usepackage{siunitx}

\usepackage{lineno}

\usepackage{array}
\newcommand{\PreserveBackslash}[1]{\let\temp=\\#1\let\\=\temp}
\newcolumntype{C}[1]{>{\PreserveBackslash\centering}p{#1}}
\newcolumntype{R}[1]{>{\PreserveBackslash\raggedleft}p{#1}}
\newcolumntype{L}[1]{>{\PreserveBackslash\raggedright}p{#1}}

\begin{document}

\title{Single-Shot Readout of a Nuclear Spin in Silicon Carbide}

\author{Xiao-Yi Lai$^{1,\,2,\,3}$}
\author{Ren-Zhou Fang$^{1,\,2,\,3}$}
\author{Tao Li$^{1,\,2,\,3}$}
\author{Ren-Zhu Su$^{1,\,2,\,3}$}
\author{Jia Huang$^{4}$}
\author{Hao Li$^{4}$}
\author{Li-Xing You$^{4}$}
\author{Xiao-Hui Bao$^{1,\,2,\,3}$}
\author{Jian-Wei Pan$^{1,\,2,\,3}$}
\affiliation{$^1$Hefei National Research Center for Physical Sciences at the Microscale and School of Physical Sciences, University of Science and Technology of China, Hefei 230026, China}
\affiliation{$^2$CAS Center for Excellence in Quantum Information and Quantum Physics, University of Science and Technology of China, Hefei, 230026, China}
\affiliation{$^3$Hefei National Laboratory, University of Science and Technology of China, Hefei 230088, China}
\affiliation{$^4$National Key Laboratory of Materials for Integrated Circuits, Shanghai Institute of Microsystem and Information Technology, Chinese Academy of Sciences, 865 Changning Road, Shanghai 200050, China}

\begin{abstract}
	Solid-state qubits with a photonic interface is very promising for quantum networks. Color centers in silicon carbide have shown excellent optical and spin coherence, even when integrated with membranes and nano-structures. Additionally, nuclear spins coupled with electron spins can serve as long-lived quantum memories. Pioneering work in previous has realized the initialization of a single nuclear spin and demonstrated its entanglement with an electron spin. In this paper, we report the first realization of single-shot readout for a nuclear spin in SiC. We obtain a deterministic readout fidelity of 98.2\% with a measurement duration of 1.13~ms. With a dual-step readout scheme, we obtain a readout fidelity as high as 99.5\% with a success efficiency of 89.8\%. Our work complements the experimental toolbox of harnessing both electron and nuclear spins in SiC for future quantum networks. 
\end{abstract}

\maketitle

Solid state color center is a promising system to demonstrate quantum computing and quantum information process~\cite{AwschalomZhou2018}. One of the most studied solid state systems is negatively charged nitrogen-vacancy (NV) centers in diamond, which have been used widely in quantum network and quantum sensing~\cite{RufHanson2021}. However, there still lacks mature growing and nanofabrication method for diamond crystals, which strictly obstacles the large scale quantum applications. The color centers in silicon carbide (SiC) are promising candidates~\cite{AtatureWrachtrup2018,WolfowiczAwschalom2021,SonAwschalom2020} with the adding of material advantage including wafer-scaling~\cite{LukinVuckovic2020} and mature fabrication~\cite{LukinVuckovic2020a}. Rapid developments with color centers in SiC have been shown during these years, including high fidelity spin and optical control~\cite{NagyWrachtrup2019,banks2019resonant}, millisecond electron spin coherence time in isotopic purified material~\cite{bourassa2020entanglement}, two photon interference~\cite{MoriokaKaiser2020}, efficiency spin-photon interface with high coherence~\cite{BabinWrachtrup2021}, single-shot readout by spin-charge conversion~\cite{AndersonAwschalom2022}, and spin-photon entanglement~\cite{fang2023experimental}. Typically, the V2 centers in SiC have shown excellent properties, e.g. high optical coherence when temperature is up to 20~K~\cite{udvarhelyi2020vibronic}, higher quantum efficiency compare to V1 center in SiC~\cite{liu2023silicon}, and can maintain narrow optical linewidth even when created by ion implantation or integrated in waveguides~\cite{BabinWrachtrup2021} and microcavities~\cite{lukin2022optical} and sub-micro membranes~\cite{heiler2023spectral}. However, due to large gyromagnetic ratio, electron spins couple strongly with the crystal local environment, which induce the the coherence time of electron spins are strictly limited by impurities and nuclear baths in the crystal~\cite{bulancea2023isotope}. 

Nuclear spins are crucial resources for quantum computation and quantum information~\cite{reiserer2016robust}. With less coupling with the local crystal environment compare with electron spins, nuclear spins usually possess long coherence time, make them perfect quantum memories~\cite{stas2022robust}. Nuclear spins in SiC have rapid developments, many milestone experiments have been demonstrated, such as entanglement between electron-nuclear spin ensembles at room temperature~\cite{klimov2015quantum}, single nuclear spin initialization~\cite{falk2015optical}, entanglement between a single divacancy and a strongly coupled nuclear spin~\cite{bourassa2020entanglement}. However, the single-shot readout of a nuclear spin is still yet to be realized.  

\begin{figure*}[htb]
	\centering
	\includegraphics[width=1\textwidth]{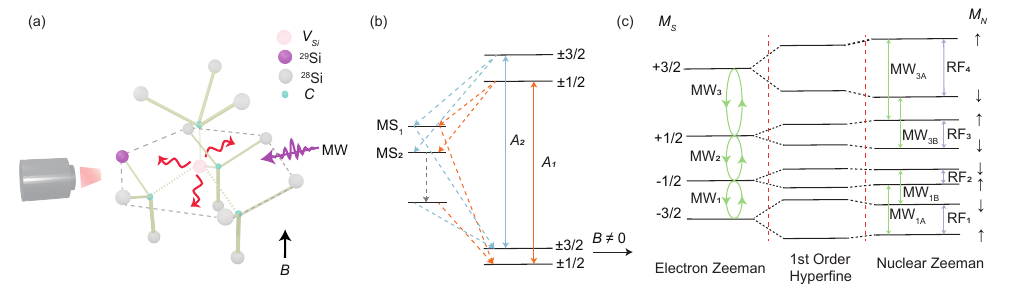}
	\caption{Energy level of a V2 center strongly coupled with a NNN ${ }^{29} \mathrm{Si}$ nuclear spin. (a) Atomic model of the negatively charged silicon-vacancy defect in 4H-SiC. In plane NNN ${ }^{29} \mathrm{Si}$ nucleus are connected by dotted lines for marking. 4H-SiC c-axis is parallel with the external magnetic field, and the fluorescence from V2 center is collected perpendicular to c-axis.
(b) Energy level diagram of V2 centres at zero external magnetic field. (c) Ground states level structure of a V2 center strongly coupled with a single nuclear spin under external magnetic field. }
	\label{fig.1}
\end{figure*}

In this paper report the first realization of single-shot readout of a nuclear spin in SiC. We choose a next-nearest-neighbor (NNN) ${ }^{29} \mathrm{Si}$ nuclear spin which is strongly coupled with a k-site silicon vacancy (V2) center (Fig.~\ref{fig.1}a). The V2 center in commercial natural abundance 4H-SiC is created by electron irradiation and post annealing, we fabricate SIL and coat $\mathrm{Al}_2 \mathrm{O}_3$ film in the a-side to enhance collection efficiency. The sample is placed into a 4~K cryostat, a confocal system is established to excite V2 and collect fluorescence from V2. The objective is outside the cryostat, with N.A. = 0.65. The zero phonon line (ZPL) wavelength of V2 centers in 4H-SiC is 916~nm, we collect phonon side band (PSB) fluorescence above 930~nm to filter resonant excite laser noise. The external magnetic field is 942~G, the main component of external magnetic field is provided by two permanent magnets installed on both sides of the sample stage, and the small angle misalignment compare to c-axis is compensated by a permanent magnet outside the cryostat. Under this external magnetic field, the V2 center electron Zeemann splitting is much larger than electron-nuclear hyperfine splitting, the flip-flop process between electron and nuclear spins is strongly suppressed, which is the basis of our single-shot readout process. There are two types of NNN ${ }^{29} \mathrm{Si}$ nuclear locations, one is that the nuclear spin and V2 center are in the same plane perpendicular to crystal c-axis, and the other is out of the plane, as shown in Fig.~\ref{fig.1}a. To further suppress flip-flop process between V2 center and nuclear spin, we choose the NNN ${ }^{29} \mathrm{Si}$ nuclear spin in the first location (purple atom in Fig.~\ref{fig.1}a) to demonstrate the experiment in the following. To demonstrate nuclear spin single-shot readout, we first initialize the V2 electron spin, then the nuclear spin is initialized by applying swap gate between electron and nuclear spin~\cite{waldherr2014quantum}, finally, the single-shot readout of nuclear spin is based on 250 readout cycles, each readout cycle contains two controlled not (CNOT) operations which map a specific nuclear spin state onto the electron spin~\cite{jiang2009repetitive}, and an electron readout pulse which can selectively readout the electron states.

According to recent work about the intrinsic spin dynamics of V2 center, we can simply treat the energy level structure of V2 center as a 6-level system~\cite{liu2023silicon} when the external magnetic field is zero, as shown in Fig.~\ref{fig.1}b. We assign the transition from $\ket{\pm 1/2}$ excited states to $\ket{\pm 1/2}$ ground states as A1 transition, and assign the transition from $\ket{\pm 3/2}$ excited states to $\ket{\pm 3/2}$ ground states as A2 transition. The ground states level structure of a V2 strongly coupled with a single ${ }^{29} \mathrm{Si}$ nuclear spin under external magnetic field is shown in Fig.~\ref{fig.1}c. In the following, for V2 center electron spin, we denote the four states as $\ket{+3/2}$, $\ket{+1/2}$, $\ket{-1/2}$, $\ket{-3/2}$, and for ${ }^{29} \mathrm{Si}$ nuclear spin, we denote $\ket{+1/2}$ as $\ket{\uparrow}$ and denote $\ket{-1/2}$ as $\ket{\downarrow}$.

\begin{figure}[htb]
	\centering
	\includegraphics[width=\columnwidth]{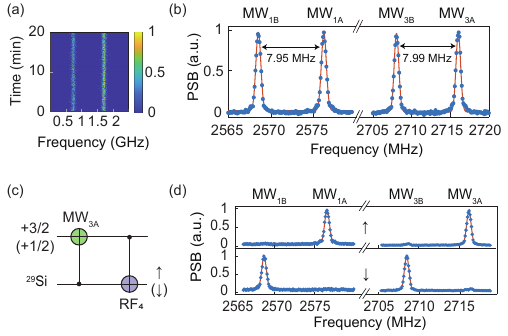}
	\caption{Electron spin properties and nuclear spin initialization. (a) Resonant excitation scans over 20 min. (b) ODMR signal of ground states under under 942~G external magnetic field, the hyperfine splitting is about 8~MHz. (c) Swap gate between nuclear spin and V2 electron spin. The electron spin is initialized in $\ket{+ 3/2}$ ($\ket{+ 1/2}$), the nuclear spin is swap to $\uparrow$ ($\downarrow$) after applying the swap gate. (d) ODMR signal of ground states after nuclear initialization, one of the two peaks separated by 8~MHz is disappear. 
}
	\label{fig.2}
\end{figure}

The optical and spin coherence of V2 center perform well in our system. We sweep the wavelength of weak continuous resonant excitation laser around 916.4~nm for 20~min, and collect PSB to determine the resonant frequency, the resonant frequency is very stable, as shown in Fig.~\ref{fig.2}a. After determining the frequencies of A1 and A2 transition, we apply an 1~ns optical resonant pulse to measure the lifetime of both transitions, which are 6.45~ns and 10.58~ns respectively. The lifetime of A2 transition is much longer than lifetime of A1 transition, this means higher quantum efficiency of A2 transition. To measure the V2 electron ODMR spectral, we sweep the MW frequency with step of 0.1~MHz. As we can see in Fig.~\ref{fig.2}b, the hyperfine splitting is about 8~MHz. The Gaussian fitting linewidth is about 0.6~MHz, which is narrow enough compare with the hyperfine coupling split. The fidelity of MW3A $\pi$-pulse is about 96.7$\pm$0.4~\%, which is limited by the short T2* coherence time ($\approx$ 0.8~$\upmu$s). Nuclear spin shows long coherent time in our commercial 4H-SiC crystal without isotope engineering. The T2* coherence time of 9.9$\pm$1.2~ms is characterized by Ramsey sequence. The T2* coherence time of nuclear spin is 4 orders longer than V2 electron spin, confirm that the nuclear spins is almost perfect quantum memories even in commercial SiC crystals. The nuclear spin T1 time is far more longer than 1~s, so the influence of nuclear spin relaxation can be neglected in our single-shot readout process.

Nuclear spin initialization is an essential step for nuclear single-shot readout measurement and for nuclear-photon entanglement. We first initialize the V2 spin to $\ket{+3/2}$ ($\ket{+1/2}$) state by applying A1 (A2) laser combining with MW1A ans MW1B for 50~$\upmu$s with 99~\% fidelity. Next, we apply a swap gate (Fig.~\ref{fig.2}c) to copy the electron spin state to nuclear spin state, the nuclear spin is initialized in $\ket{\uparrow}$ ($\ket{\downarrow}$). To test the fidelity of nuclear spin initialization, we measure V2 center electron spin ODMR signal after swap gate. As we can see in Fig.~\ref{fig.2}d, one of the two peaks separated by 8~MHz is disappear after the nuclear spin is initialized in $\ket{\uparrow}$ ($\ket{\downarrow}$), we can conclude that the fidelity of nuclear initialization is about 93$\pm$0.5~\%. 

\begin{figure}[htb]
	\centering
	\includegraphics[width=\columnwidth]{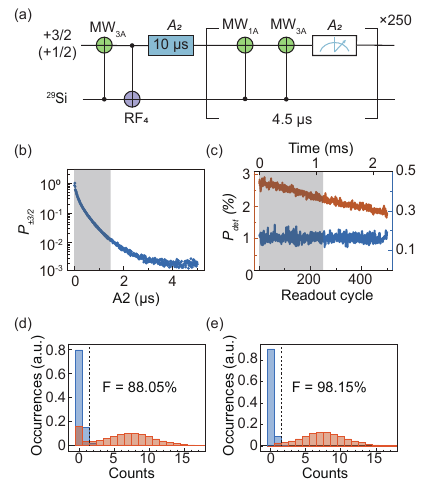}
	\caption{Nuclear spin single-shot readout.  (a) Representation of the nuclear spin single-shot readout scheme.  (b) The photoluminescence signal during 5~$\upmu$s A2 laser excitation. The optical pumping fidelity of the first 1.5~$\upmu$s reaches over 98.5~\% (gray shaded area). (c) Detection probability of 1.5~$\upmu$s A2 laser during 500 readout cycles after nuclear spin initialized in bright state (yellow) and dark state (blue). In the following single-shot readout experiment, we only readout for 250 cycles (gray shaded area). (d) Photon-count distribution for nuclear spin single-shot readout after initialized in bright state (yellow bar) and dark state (blue bar). The raw fidelity is 88.05$\pm$0.14~\%, which contains nuclear spin initialization and readout. The average photons detected from bright (dark) state is 6.24 (0.4). (e), The conditional photon-count distribution, shows the average single-shot readout fidelity of bright and dark state achieves 98.15$\pm$0.07~\%.
 }
	\label{fig.3}
\end{figure}

We now proceed to single-shot readout of NNN ${ }^{29} \mathrm{Si}$ nuclear spin. As depicted in Fig.~\ref{fig.3}a, we initialize the nuclear spin in $\ket{\uparrow}$ or $\ket{\downarrow}$ state, and then apply 10~$\upmu$s weak continuous A2 laser to pump the electron spin population to $\ket{\pm 1/2}$ states. Finally, we readout the nuclear spin state for 250 readout cycles. During each cycle, we first apply two CNOT operations (MW1A and MW3A $\pi$-pulse) on V2 electron spin, and then apply 1.5~$\upmu$s A2 readout laser. The CNOT gates will flip the V2 electron spin population back to $\ket{\pm 3/2}$ state only if the nuclear spin is in $\ket{\uparrow}$ state, so that the V2 center can emit photons in the next readout cycle. According to the readout sequence, we can denote $\ket{\uparrow}$ state of the nuclear spin as bright state, and denote $\ket{\downarrow}$ state as dark sate. 

Before single-shot readout, we first investigate the electron-nuclear flip-flop process during 500 readout cycles, as shown in Fig.~\ref{fig.3}c. Ideally, if the nuclear spin is initialized in bright state (Fig.~\ref{fig.3}c, yellow), the photon detection efficiency of each readout cycle among 500 readout cycles is almost the same, which should equals to 2.8~\%. However, the detection efficiency decays from the first to the last cycle, this means there is a small possibility for each readout cycle that the nuclear spin flips from bright state to dark state. From this result, we deduce the nuclear flipping possibility of  \num{7.7e-4} 
 for each readout cycle.

Next, we perform nuclear spin single-shot readout experiment, each single-shot readout process contains once nuclear initialization and 250 readout cycles, the total duration of each readout experiment is 1.24~ms. We count the number of photons detected in each single-shot readout process and create a histogram, as depicted in Fig.~\ref{fig.3}d. The nuclear spin is initialized in bright state (yellow bar) or dark state (blue bar). We set N = 1 as the cutoff, corresponding to a false-negative rate $p(\uparrow\mid \downarrow)$= 0.191 and false-positive rate $p(\downarrow\mid \uparrow)$  = 0.048. $p(\uparrow\mid \downarrow)$ means the nuclear state is in bright state ($\ket{\uparrow}$) but single-shot readout result refers to dark state ($\ket{\downarrow}$), and $p(\downarrow\mid \uparrow)$ means the nuclear state is in dark state ($\ket{\downarrow}$) but single-shot readout result refers to bright state ($\ket{\uparrow}$). The overall fidelity of initialization and single-shot readout is 88.05$\pm$0.14~\%. 

To exclude the infidelity from other process such as nuclear initialization, we set a condition for the readout data: for bright state, if we detect at least 1 photon in the first 120 readout cycles, we save the 250 cycles results of this single-shot readout process for analysis, and for dark state, if we don't detect a photon in the first 120 readout cycles, we save the 250 cycles result for analysis. This conditional single-shot readout results are shown in Fig.~\ref{fig.3}e. This conditional results mitigates the effect of initialization error and charge state error, which will severely restrict the overall single-shot readout fidelity. Finally, we achieve an average single-shot readout fidelity of 98.15$\pm$0.07~\%, which contains the false-negative rate $p(\uparrow\mid \downarrow)$= 0.028 and the false-positive rate $p(\downarrow\mid \uparrow)$  = 0.009. 

\begin{figure}[htb]
	\centering
	\includegraphics[width=\columnwidth]{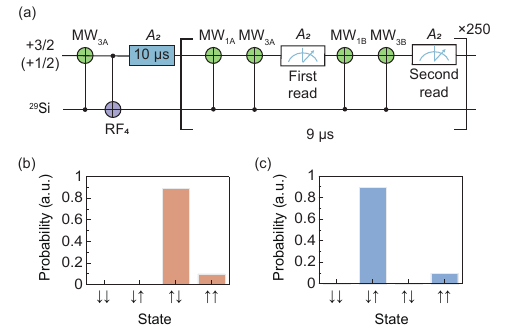}
	\caption{Nuclear spin dual-step readout. (a) Circuit diagram of the scheme. (b) and (c) The joint statistics results of the first and the second read after initialized in bright state and dark state. By considering the $\ket{\uparrow \downarrow}$ and $\ket{\downarrow \uparrow}$ subspace, the single-shot readout fidelity reaches 99.52$\pm$0.09~\% (b) and 99.50$\pm$0.09~\% (c).
 }
	\label{fig.4}
\end{figure}

Furthermore, by sacrificing the readout success probability, we can achieve higher single-shot readout fidelity with dual-step single-shot readout scheme~\cite{kindem2020control}, as depicted in Fig.~\ref{fig.4}a. The initialization process is the same as above, and a single-shot readout process is also consists of 250 readout cycles. During each readout cycle, we first apply two CNOT gates (MW1A and MW3A $\pi$-pulse) on V2 electron spin and 1.5~$\upmu$s A2 readout laser for the first read, then we apply another two CNOT gates (MW1B and MW3B $\pi$-pulse) on V2 electron spin and 1.5~$\upmu$s A2 readout laser for the second read.

The joint statistics results of two successive reads is shown in Fig.~\ref{fig.4}b-c. We assign the nuclear spin into the bright state ($\ket{\uparrow}$) if we measure two or more photons during the first readout sequence and one or zero photons during the second readout sequence, and vice versa for the dark state ($\ket{\downarrow}$). In Fig.~\ref{fig.4}b, the nuclear is initialized in bright state, the joint statistics results show large $\ket{\uparrow \uparrow}$ probability which comes from nuclear flip. If we only consider the $\ket{\uparrow \downarrow}$ and $\ket{\downarrow \uparrow}$ subspace, we can achieve single-shot readout fidelity of 99.52$\pm$0.09~\% with a success efficiency of 89.77~\%. Also, after nuclear initialization in dark state (Fig.~\ref{fig.4}c), the single-shot readout fidelity is 99.50$\pm$0.09~\% if only consider the $\ket{\uparrow \downarrow}$ and $\ket{\downarrow \uparrow}$ subspace, the success efficiency is 89.82~\%. 

Based on our single-shot readout scheme, decreasing the nuclear flip-flop probability and improving the collection efficiency will improve the readout fidelity. Firstly, to decrease the V2 center-nuclear flip-flop possibility, 4H-SiC with isotope engineering is crucial~\cite{ivanov2014high}, which have been used in previous works and exhibit excellent coherence properties with V2 center Hahn-echo coherence time reach 1.3~ms~\cite{BabinWrachtrup2021}. By using this sample, ultra-narrow V2 center ODMR linewidth around 100~kHz can be achieved. When this V2 center is coupled with a ${ }^{29} \mathrm{Si}$ nuclear spin which is a little farther than NNN ${ }^{29} \mathrm{Si}$, the splitting of V2 center energy levels is about 2~MHz~\cite{hesselmeier2023measuring}. Under this energy splitting, two narrow V2 center ODMR peaks can still be manipulated independently. Smaller splitting means less coupling between V2 center and nuclear spin, which can significantly reduce the possibility of flip-flop process during each readout cycle, and as a result, the nuclear single-shot readout fidelity can be almost perfect. 

Secondly, to improve the collection efficiency, we can couple the V2 center to nanostructures~\cite{majety2022quantum}, which have been demonstrated by many works before. The overall collection efficiency can be highly enhanced by at least 5 times after coupling with nanostructures such as photonic crystal cavities~\cite{lukin20204h} or microdisk cavities~\cite{lukin2022optical} and collected by taper fibers~\cite{nguyen2019quantum} or grating couplers~\cite{dory2019inverse}. As a result, the single-shot readout fidelity will achieve 99.7~\% and the whole time for single-shot readout will be shorten from 1.1~ms to 0.2~ms. By combine crystal isotope engineering and cavity enhancement together, nuclear single-shot readout with high speed and perfect fidelity is expected in the future.

In conclusion, we perform the first demonstration of nuclear spin single-shot readout in silicon carbide. The high readout fidelity (99.5$\pm$0.09~\%) demonstrated in this work creates opportunities for high fidelity nuclear photon entanglement with long lifetime. Even though the optical pumping process includes complex dynamics in the V2 center, the nuclear flip-flop process is slow enough to perform single-shot readout by using the in plane NNN ${ }^{29} \mathrm{Si}$ nuclear spin. The coherence properties of V2 electron spins and nuclear spins can be improved by crystal engineering, such as isotope purification and decreasing of impurity concentration. Thanks to weak coupling to the stray electric field and strain, V2 center in silicon carbide can be integrated into nanophotonic structures, such as 1D photonic crystal cavity and micro-disk cavity. When the cavity is strongly coupled, the single-shot readout speed and fidelity will be greatly improved. The ZPL wavelength of the V2 centers can be converted to telecom U band with a pump laser near 2051~nm, allowing long-distance transmission in optical fibers~\cite{2020Entanglement}. With these improvements, the V2 silicon vacancy in SiC may become a highly competitive approach for quantum networks~\cite{WehnerHanson2018}.

\section{Acknowledgment}
The nanofabrication was carried out at the USTC Center for Micro- and Nanoscale Research and Fabrication. This research was supported by the Innovation Program for Quantum Science and Technology (No. 2021ZD0301103), National Natural Science Foundation of China, and the Chinese Academy of Sciences.

\bibliography{main}

\end{document}